\documentclass[12pt,a4paper]{article}
\pdfoutput=1
\usepackage{graphicx}
\usepackage{appendix}
\usepackage{amsmath,amsfonts,amssymb,setspace}
\usepackage{braket,xcolor}
\usepackage{caption}
\usepackage{subcaption}
\usepackage{hyperref}
\usepackage{cite}
\usepackage{tocloft}
\usepackage[margin=0.75in]{geometry}
\hypersetup{colorlinks=false, linkcolor=blue, citecolor=red}
	\def\be{\begin{equation}}
	\def\ee{\end{equation}}

	\newcommand{\vs}[1]{\vspace{#1 mm}}
	
\begin{document}
	\begin{flushright}
		%
	\end{flushright}
	\begin{center}
		{\Large{\bf Complexity for Open Quantum System}}\\

\vs{10}

{\large
Arpan Bhattacharyya${}^{a,\,}$\footnote{\url{abhattacharyya@iitgn.ac.in
}},Tanvir Hanif ${}^{b,\,}$\footnote{\url{thanif@du.ac.bd}}, S. Shajidul Haque${}^{c,d,\,}$\footnote{\url{shajid.haque@uct.ac.za}},\\ Md. Khaledur Rahman ${}^{b,\,}$\footnote{\url{mdkhaledur-2017812513@tphy.du.ac.bd}}}

\vskip 0.3in

{\it ${}^{a}$ Indian Institute of Technology, Gandhinagar, Gujarat-382355, India}\vskip .5mm

{\it ${}^{b}$ Department of Theoretical Physics, University of Dhaka, Dhaka, Bangladesh}\vskip .5mm

{\it ${}^{c}$
High Energy Physics, Cosmology \& Astrophysics Theory Group \\ and \\ The Laboratory for Quantum Gravity \& Strings \\ Department of Mathematics and Applied Mathematics, \\ University of Cape Town, South Africa }\vskip .5mm

{\it ${}^{d}$ National Institute for Theoretical and Computational Sciences (NITheCS)\\ South Africa}

\vskip.5mm

\end{center}

\vskip 0.35in

\begin{abstract}
We study the complexity for an open quantum system. Our system is a harmonic oscillator coupled to a one-dimensional massless scalar field, which acts as the bath. Specifically, we consider the reduced density matrix by tracing out the bath degrees of freedom for both regular and inverted oscillator and computed the complexity of purification (COP) and complexity by using the operator-state mapping. We found that when the oscillator is regular the COP saturates quickly for both underdamped and overdamped oscillators. Interestingly, when the oscillator is underdamped, we discover a kink like behaviour for the saturation value of COP with varying damping coefficient. For the inverted oscillator, we found a linear growth of COP with time for all values of bath-system interaction. However, when the interaction is increased the slope of the linear growth decreases, implying that the unstable nature of the system can be regulated by the bath. 
	\end{abstract}

	\newpage

\tableofcontents



\section{Introduction}
We consider the circuit complexity of an open quantum system, i.e. a system not in isolation but coupled to a ``bath". Motivation for such a system stems from the fact that most experimentally accessible systems are open quantum systems \cite{CALDEIRA1983374, PhysRevLett.46.211}. Our model consists of a harmonic oscillator as the system, and a one dimensional bosonic (free) field theory as the bath \cite{zurekbath}. This model is exactly solvable and provides a perfect setup to explore the effect of the bath on a quantum system. We will consider a new approach, namely, the complexity of purification, to study this influence. Apart from considering the system as a regular harmonic oscillator, we will also consider it as an inverted harmonic oscillator. An inverted harmonic oscillator has an unstable fixed point and is well known as the benchmark toy model to explore chaotic like behaviour \cite{Ali:2019zcj}. For inverted oscillator, our goal is to understand how a bath can affect such a system's unstable (chaotic like) behaviour. 

Complexity, a tool of quantum information theory, measures the shortest distance between some reference states $|\psi\rangle_R$ and a target state $|\psi\rangle_T$. Operationally, it quantifies the minimal number of operations needed to manipulate $|\psi\rangle_R$ to $|\psi\rangle_T$. The flurry of recent work on circuit complexity in the field of quantum field theory in recent time \cite{Jefferson,Chapman:2017rqy,Caputa:2017yrh,me1,Bhattacharyya:2018bbv,Hackl:2018ptj,Khan:2018rzm,Camargo:2018eof,Ali:2018aon,Bhattacharyya:2018wym,Caputa:2018kdj,Bhattacharyya:2019kvj,Caputa:2020mgb,Flory:2020eot,Erdmenger:2020sup, cosmology1,cosmology2,DiGiulio:2020hlz,Caceres:2019pgf,Susskind:2020gnl,Chen:2020nlj,Czech:2017ryf,Camargo:2019isp,Chapman:2018hou,Chapman:2019clq,Doroudiani:2019llj,Geng:2019yxo,Guo:2020dsi,Haque:2021hyw,Haque:2021kdm} \footnote{This list is by no means exhaustive. Interested readers are referred to these reviews \cite{Chapman:2021jbh, Bhattacharyya:2021cwf}, and citations are therein for more details.} has largely been spurred by black hole physics and, in particular, the conjecture that it resolves certain puzzles related to black holes \cite{susskind1, susskind2}. 

In this paper, we will consider an open system represented by the reduced density matrix obtained by tracing out the bath. Then we will purify the state and compute complexity of purification (COP) \cite{Agn2019,Caceres:2019pgf, Camargo:2018eof, Camargo:2020yfv}. We will also consider complexity by using the state operator mapping, which essentially picks out a particular purification and lacks the scanning for the minimum path. Interestingly, we will see that, for our open system, operator-state mapping is also less sensitive in terms of capturing the underlying dynamics of the system. 

Indeed one might be tempted to compute \textit{operator complexity} \cite{Balasubramanian:2019wgd} for the reduced density operator. However, it is not a unitary operator. For operator complexity, usually, one computes the optimal circuit, which connects identity with the corresponding operator. The optimal circuit that we construct is a unitary operator itself even when we consider wavefunctions. For instance, when one computes operator complexity , one typically deals with a time evolution operator (for this reason sometimes it is known as the Hamiltonian complexity), which is a unitary operator. In contrast, we do not have a \textit{unitary circuit} that connects identity with the reduced density matrix since the reduced density matrix is not a unitary operator. Therefore, COP and the complexity by operator-state mapping are the most natural approaches for realizing complexity for the reduced density matrix. 

One particular aspect of complexity that we are interested in is the effect of the bath on the system during the transition of the oscillator from underdamped to overdamped regime by changing the amount of damping. 
In addition, for inverted oscillator our interest is to check if complexity can be used to regulate the unstable (chaotic-like) behaviour. Previous works \cite{Ali:2019zcj,Bhattacharyya:2019txx,Ryuchaos,Bhattacharyya:2020art,Balasubramanian:2019wgd,Yang:2019iav,Yang:2019udi} showed that complexity can detect the scrambling time and Lyapunov exponent that emerge from a single inverted harmonic oscillator. 
In this paper, we would like to explore how complexity behaves when the instability of the inverted harmonic oscillator is regulated by the bath, for which we will consider a string. At the operational level, we compute the complexity of purification of the reduced density matrix by tracing out the string. The interaction term, which plays the role of damping, will affect the evolution of the complexity of purification. 

The paper is organized as follows.Section \ref{sec:sp_purifcn}, has a quick review on the complexity of purification and the state operator mapping. Section \ref{sec:model} gives the details of our model. In section \ref{sec:COP} we compute and discuss the results of the complexity of purification. In section \ref{sec:state} we discuss circuit complexity from the state operator mapping. In Section \ref{sec:Discussion}, we conclude with a discussion of our results and future directions.


\section{Complexity for a Density Matrix}
\label{sec:sp_purifcn}
The density matrix for a mixed state $\hat \rho_{\rm mix}$ on the Hilbert space $\cal H$, can be purified to a pure state $|\Psi\rangle$ in an enlarged Hilbert space ${\cal H}_{\rm pure} = \cal H \otimes \cal H_{\rm anc}$, where ${\cal H}_{\rm anc}$ corresponds to an ``ancillary'' set of degrees of freedom. 
The key point of this purification is that the trace of the density matrix of $|\Psi\rangle$ over the ancillary degrees of freedom must generate the original mixed state 
${\rm Tr}_{\rm anc}\left(|\Psi\rangle\langle \Psi|\right) = \hat \rho_{\rm mix}$. 
Note that expectation values of operators acting in ${\cal H}$ are preserved under purification, $\langle \hat {\cal O}\rangle = {\rm Tr}_{\rm anc}\left(\langle \Psi|\hat {\cal O}|\Psi\rangle\right) = {\rm Tr}\left(\hat \rho_{\rm mix}\hat {\cal O}\right)$, so that observables are preserved by purification. 

Due to the arbitrariness of the choice of the ancillary Hilbert space ${\cal H}_{\rm anc}$ the purification process is not unique. 
For example, there may be a set of pure states $\left\{|\Psi\rangle_{\alpha,\beta,...}\right\}$, parameterized by $\alpha, \beta, ...$, all of which satisfy the purification requirement.
By minimizing a quantity of interest, such as the entanglement entropy or complexity, with respect to the parameters it is possible to distinguish among different purifications. In this work, we are investigating the complexity of the mixed state, therefore, we will minimize the complexity of the set of purifications $\left\{|\Psi\rangle_{\alpha,\beta,...}\right\}$ of $\hat \rho_{\rm mix}$, obtaining the {\bf complexity of purification} (COP) \cite{Agn2019,Caceres:2019pgf,Camargo:2018eof,Camargo:2020yfv},
\be
{\cal C}_{\rho} = \min_{\alpha, \beta,...} {\cal C}\left(|\Psi\rangle_{\alpha, \beta,...},|\psi_R\rangle\right)\, ,
\label{eq:CoP}
\ee
where we made explicit the dependence of the complexity of the pure state on the reference state $|\psi_R\rangle$.


An alternative approach to purify the reduced density matrix $\hat \rho_{\rm mix}$ is by using the technique of {\bf operator-state mapping} (also known as the channel-state mapping) \cite{CHOI1975285,JAMIOLKOWSKI1972275}.
To implement this technique, one need to consider an operator on the Hilbert space ${\cal H}$ with representation $\hat {\mathcal O} = \sum_{m,n} {\mathcal O}_{mn} |n\rangle\langle m|$. In its simplest form, the mapping associates a state $|{\mathcal O}\rangle$ to $\hat {\mathcal O}$ by flipping the bra to a ket,
\be
\hat {\mathcal O} = \sum_{m,n} {\mathcal O}_{mn} |n\rangle\langle m| \hspace{.2in} \longleftrightarrow \hspace{.2in}|{\mathcal O}\rangle = \frac{1}{\sqrt{\text{Tr} [\mathcal{O}^\dagger \mathcal{O}]}} \sum_{m,n} \mathcal{O}_{mn} | m \rangle \otimes | n \rangle_{\rm anc}\, .
\label{eq:OSOperatorMapping}
\ee
The state $|{\mathcal O}\rangle$ exists on the doubled Hilbert space ${\mathcal H}\otimes {\mathcal H}_{\rm anc}$. Once again we have denoted the extra copy of ${\mathcal H}$ as ${\mathcal H}_{\rm anc}$ to distinguish it from the original.
In that sense, this process of operator-state mapping is a special case of the purification discussed above \cite{haqueReducedDensityMatrix}. The most important difference is that the state $|\mathcal{O}\rangle$ in (\ref{eq:OSOperatorMapping}) associated to the operator $\hat {\mathcal O}$ is unique, there are no free parameters introduced in the mapping. Hence, the complexity associated with the operator-state mapping does not require a minimization.

\section{Our Model}
\label{sec:model}
The system we consider is a harmonic oscillator coupled to a one-dimensional bosonic (free) field theory \cite{zurekbath}. The Hamiltonian is
\begin{equation}
 H = \int_0^L \hspace{-0.1in} dx~ \left\{ 
    \frac{1}{2} \left( \Pi^2 + (\partial_x \phi)^2 \right)
 + \delta(x) \left( \frac{1}{2} P^2 + \frac{\omega^2_0}{2} Q^2 
      + \lambda Q \partial_x \phi  \right)  \right\} \ ,
\label{Hcanonical}
\end{equation}
where we are ultimately interested in the limit $L \rightarrow \infty$.
In Eq.~\ref{Hcanonical}, $Q$ and $P$ are canonically conjugate variables describing the system, $[Q,P] = i$ and
$\phi$ and $\Pi$ are canonically conjugate fields of the bath, 
$[ \phi(x), \Pi(x') ] = i \delta(x - x')$.\footnote{We work in units where $\hbar=1$.  
We have also set the string's speed of sound to unity: $v=1$.}
Furthermore, we will consider that the field $\phi$ satisfies the Dirichlet boundary conditions:
\begin{equation}
\phi(x = 0) = 0,  \ \ \ \phi(x = L)= 0
\label{boundary}
\end{equation}
We consider a quench in the above model (Eq.~\ref{Hcanonical}) and we have the following quench protocol
\begin{equation}
 H = \left\{ \begin{array}{c c c}
    H_<  & {\rm for}  &  t < 0  \\
    H_>  & {\rm for}  &  t > 0
    \end{array} \right.  \ ,
\end{equation}
In what follows, we start the system in the ground state of $H_<$;
we evolve the system with $H_>$.
We take $H_<$ to be the Hamiltonian when the system and bath are decoupled
\begin{equation}
 H_< = \int_0^L \hspace{-0.1in} dx~
   \left \{ \frac{1}{2} \left ( \Pi^2 + (\partial_x \phi)^2 \right)
 + \frac{1}{2} \left ( P^2 + \Omega_<^2 Q^2 \right) \right\} \ .
\label{Hinit}
\end{equation}
We quench in the system-bath coupling $\lambda$ and the system's frequency so that $H_>$ is given by Eq.~\ref{Hcanonical}.
Now performing the following mode expansions for the system and the bath
\begin{eqnarray}
Q &=& \frac{1}{\sqrt{2\Omega_<} } ( a_0 + a_0^{\dagger} ) \\
 P &=& -i \sqrt{ \frac{\Omega_<}{2} } ( a_0 - a_0^{\dagger} ) \\ 
 \phi(x)& = & \sqrt{\frac{2}{L}} \sum_{n=1}^{\infty} \sin \left( \frac{\pi n}{L} x \right)
 \frac{1}{\sqrt{2\Omega_n} } ( a^{\phantom \dagger}_n + a^{\dagger}_n ) \\
 \Pi(x) &=& -i \sqrt{\frac{2}{L}} \sum_{n=1}^{\infty} \sin \left( \frac{\pi n}{L} x \right)
 \sqrt{\frac{\Omega_n}{2} } ( a^{\phantom \dagger}_n - a^{\dagger}_n )  \ ,
\label{modedirichlet}
\end{eqnarray}
one obtains the following decoupled Hamiltonian
\begin{equation}
 H_< = \sum_{n=1}^{\infty} \Omega_n \left( a^{\dagger}_n a^{\phantom \dagger}_n + \frac{1}{2} \right)
 + \Omega_< \left( a_0^{\dagger} a_0^{\phantom \dagger} + \frac{1}{2} \right) \ ,
\nonumber
\end{equation}
where $\Omega_n = \frac{\pi n}{L}$. 
Our initial state $| \psi_0 \rangle$ is annihilated by the $\{ a_i \}$; 
our final state is obtained by evolving $| \psi_0 \rangle$ with $H_>$:
\begin{equation}
 a_i \mid \psi_0 \rangle = 0 \ \  \forall \  i  
 \ \ ; \ \ 
 \mid \psi(t) \rangle = \exp(-i H_> t) \mid \psi_0 \rangle \ .
\end{equation}
To proceed, we introduce the dual field $\theta$ such that $\Pi = - \partial_x \theta$
with \begin{equation} [ \phi(x), \theta(x') ] =  \frac{i}{2} \ {\rm sgn}(x' - x).\end{equation}
The bath part of the Hamiltonian takes the form
\begin{equation}
 H_B = \int_0^L \hspace{-0.1in} dx~ 
    \frac{1}{2} \left[ (\partial_x \theta)^2 + (\partial_x \phi)^2 \right] \ .
\nonumber
\end{equation}
Then, we decompose $\phi$ and $\theta$ into right and left movers
\begin{subequations}
\begin{equation}
 \phi(x) = \phi_R(x) + \phi_L(x)
 \ \ , \ \ 
 \theta(x) = \phi_R(x) - \phi_L(x)\ ,
\end{equation}
where $\phi_R$ and $\phi_L$ satisfy the following commutation relations
\begin{equation}
[\phi_R(x), \phi_R(x')] = \frac{i}{4} {\rm sgn}(x-x'),\, 
[\phi_L(x), \phi_L(x')] = -\frac{i}{4} {\rm sgn}(x-x'),\,
[\phi_R(x), \phi_L(x')] = 0\,.
\end{equation} 
\end{subequations}
In the Heisenberg picture, 
$\phi_R$ is a function of $(x-t)$, and $\phi_L$ is a function of $(x+t)$:
\begin{align*}
\phi_R(x,t) = \phi_R(x-t) \ \, \ \ \phi_L(x,t) = \phi_L(x+t).
\end{align*}
%
Using the Dirichlet boundary condition at $x=0$ (\ref{boundary}) 
we have $ \phi_L(x=0,t) = - \phi_R(x=0,t)$. 
This implies that  $\phi_L$ can be regarded as the continuation of $\phi_R$ to $x < 0$:
\begin{subequations}
\begin{equation}
 \phi_L(x+t) = - \phi_R(-x-t) \ ,
\end{equation}
Similarly, using the boundary condition at $x=L$ (\ref{boundary}) 
we obtain
\begin{equation}
\phi_R(-L,t) = \phi_R(L,t) \ .
\end{equation}
\end{subequations}
This implies that we can work with right-movers on the interval $-L < x < L$ satisfying periodic boundary conditions.
Now we can rewrite the Hamiltonian solely in terms or right movers
\begin{equation}
H = \int_{-L}^{L} \hspace{-0.1in} dx~ \left\{ (\partial_x \phi_R)^2
   + \delta(x) \left[ \frac{1}{2} P^2 + \frac{\omega^2_0}{2} Q^2 
      + 2 \lambda\, Q \partial_x \phi_R  \right]  \right\} \ ;
\label{Hchiral}
\end{equation}
From this Hamiltonian, we get the following (Heisenberg) equations of motion for the operators
\begin{eqnarray}
  \partial_t \phi_R &=& - \partial_x \phi_R - \delta(x)\, \lambda\, Q\, \nonumber \\
 \frac{d^2 Q}{dt^2} &=& - \omega^2_0 Q - 2 \lambda\, \partial_x \phi_R(0)
 \ .
\label{heisenberg}
\end{eqnarray}
From the homogeneous part of the first equation in Eq.~\ref{heisenberg} it is evident that $\phi_R$ is a function of $(x-t)$. Using the fact that the ``source" term acts only at $x=0$ and integrating about a small region about $x=0$, $-\epsilon<x<+\epsilon$,
Eq.~\ref{heisenberg} becomes
\begin{eqnarray}
0 &=& \phi^+_R(x=+\epsilon) - \phi^-_R(x=-\epsilon)\Big|_{\epsilon\rightarrow 0} + \lambda\, Q \nonumber \\
 \frac{d^2 Q}{dt^2} + \omega^2_0 Q &= &
  - \lambda \left[ \partial_x \phi^+_R(x=0) + \partial_x \phi^-_R(x=0) \right]
 \ ,
\label{scattering}
\end{eqnarray} 
where we have introduced the notation
$\phi^+_R(x) \equiv \phi_R(x>0)$ and $\phi^-_R(x) \equiv \phi_R(x<0)$.


To treat Eq.~\ref{scattering}, we employ the scattering formalism \cite{buttiker} 
--- we take particles as incident from $x < 0$;
they are scattered by the system (at $x=0$), and are outgoing for $x > 0$.
Practically speaking, we solve for $Q$ and $\phi^+_R$ in terms of $\phi^-_R$. Finally we get
\begin{equation}
    \frac{d^2 Q}{dt^2} + \omega_0^2 Q  +2 \Gamma \dot Q = 2 \sqrt{2\Gamma} \dot{\phi_R}^{-},
\end{equation}
where  $\Gamma = \lambda^2/2$. 
The solution for $Q(t)$ can be written as
\begin{equation}
    Q(t)= Q_H(t)+Q_P(t),
\end{equation}
where, $Q_H(t)$ and $Q_P(t)$ are the homogeneous and particular solutions respectively. The homogeneous solution turns out to be the following:
\begin{eqnarray} \nonumber
 Q_H(t) & = & e^{-\Gamma t} \left\{ [ \cos \Omega t + \left(\frac{\Gamma}{\Omega} \right)\sin \Omega t ] Q(t=0)
        +(1/\Omega) \sin \Omega t~ P(t=0)  \right\} 
  \label{solutionQ} \\
        & - & e^{-\Gamma t} \left\{ [ \cos \Omega t + \left(\frac{\Gamma}{\Omega} \right) \sin \Omega t ] Q_P(t=0)
        + (1/\Omega) \sin \Omega t~ \dot{Q}_P (t=0) \right\}  \ ,
 \end{eqnarray} 
 where $\Omega= \sqrt{\omega_0^2-\Gamma^2}$. To find the particular solution we introduce the Fourier transforms\footnote{At this point, we have taken $L \rightarrow \infty$.}
\begin{equation}
 \phi_R(x,t) = \int \hspace{-0.069in} \frac{d\omega}{\sqrt{2\pi} } e^{-i \omega (t-x) } \phi_R(\omega)
 \ \ , \ \ 
 Q(t) = \int \hspace{-0.069in} \frac{d\omega}{\sqrt{2\pi} } e^{-i \omega t} Q(\omega)  \ ,
\end{equation}
and obtain
\begin{eqnarray}
 Q_P(\omega) & = & \left( \frac{ i 2 \sqrt{2\Gamma} \omega}
     { \omega^2 + i 2\Gamma \omega - \omega_0^2} \right) \phi_R^{-}(\omega)  \ ,  
 \nonumber  \\  
 \phi_R^{+}(\omega) &=& - \sqrt{2\Gamma} \ Q_H(\omega) 
      + \left( \frac{ \omega^2 - i 2 \Gamma \omega - \omega_0^2}
          { \omega^2 + i 2\Gamma \omega - \omega_0^2} \right) \phi_R^{-}(\omega)  \ .
\label{solutionR}
\end{eqnarray}

\noindent
To execute calculations for the bath, one must understand $\phi_R(\omega)$ as
\begin{equation}
 \phi_R(\omega) = \left\{ \begin{array}{c c c}
   \frac{a^{\phantom \dagger}_{\omega}}{\sqrt{2\omega}} & {\rm for} & \omega > 0  \\
   \frac{ a^{\dagger}_{\omega}}{\sqrt{2|\omega|}} & {\rm for} & \omega < 0  \\ 
    \end{array} \right.  \ ,
\nonumber
\end{equation}
namely, that $\phi_R(\omega)$ is an annihilation operator for $\omega > 0$ 
and a creation operator for $\omega < 0$ (with a prefactor $1/\sqrt{2|\omega|}$).
Interested readers are referred to Appendix~\ref{A} for more details.
\subsection*{Reduced Density Matrix}
We will be interested in analyzing the reduced density matrix. To form the reduced density matrix, we partition the system into two subsystems, $A$ and $B$ and trace out subsystem-$B$.\newpage For our case, we trace out the bath (string) and get the reduced density matrix for the system (oscillator). 
\begin{equation}
 \hat{\rho_S} = {\rm Tr}_B[ \hat{\rho} ] \ , 
\end{equation}
where $\hat{\rho}$ is the density matrix of the entire system. The subscript $S$ and $B$ implies system and bath respectively. 

In the position representation ---
\begin{subequations}
\begin{equation}
 \hat{\rho}_S = \int dx dx'~ \rho_S(x,x') ~ | x \rangle \langle x' |  \ ,
\label{rhosystem}
\end{equation}
where
\begin{equation}
 \rho_S(x,x') = \int \prod_i dq_i~ \rho( x,\{ q_i \} \mid x', \{q_i\} ) 
\end{equation}
with
\begin{equation}
 \rho(x,\{ q_i \} \mid x', \{q'_i \} ) = \psi(x, \{ q_i \} ) \psi^*( x', \{ q'_i \} )
\label{2particlerho}
\end{equation}
\end{subequations}
%
As the initial density matrix is Gaussian and the Hamiltonian is quadratic, the density matrix remains Gaussian for all times; in the position representation,
it has the structure
\begin{equation}
 \rho_S(x,x') = \sqrt{\frac{\gamma_1 - \eta}{\pi} }
   \exp \left \{ - \frac{\gamma}{2} x^2 -\frac{\gamma^*}{2} x'^2 + \eta x x' \right \} \ ,
\label{rhoSstructure}
\end{equation}
where $\gamma \ (= \gamma_1 + i \gamma_2) \in {\bf C}$ and $\eta \in {\bf R}$.

The parameters $\{\gamma, \eta \}$ can be obtained from the system's correlation functions. 
By definition, the reduced density matrix reproduces all correlators of the system; as the  density matrix is Gaussian, it is fully characterized by its first and second moments. The details are given in Appendix~\ref{B}.
By assumption, the initial wave function has $\langle Q \rangle =0$ and 
$\langle P \rangle = 0$; due to the structure of the Hamiltonian, this is preserved for all times. We now introduce the following notation,
\begin{subequations} \label{correlator}
\begin{equation}
 \sigma_{Q} = \langle Q(t) Q(t) \rangle  \ \ , \ \ 
 \sigma_{P} = \langle P(t) P(t) \rangle  \ \ , \ \  
 \sigma_{QP} = \frac{1}{2} \langle Q(t) P(t) + P(t) Q(t) \rangle  \ \ ;
\end{equation}
one readily obtains
\begin{equation}
 \gamma_1 = \sigma_P - \frac{1}{\sigma_Q} \left( \sigma_{QP}^2 - \frac{1}{4} \right)
 \ \ , \ \ 
 \gamma_2 = - \frac{\sigma_{QP} }{\sigma_Q}
 \ \ , \ \ 
 \eta = \sigma_P - \frac{1}{\sigma_Q} \left( \sigma_{QP}^2 + \frac{1}{4} \right)  \ .
\end{equation} 
\end{subequations}
Again, the details of the computation of these correlators are given in Appendix~\ref{B} and \ref{C}. 
\section{Complexity of Purification}
\label{sec:COP}
In this section, we will investigate the influence of the bath on the system by using the complexity of purification for the reduced density matrix, where the bath degrees of freedom are traced out. The coupling $\lambda$ and hence the damping $\Gamma$ (since $\Gamma=\lambda^2/2$) dictates the influence of the bath on the system. A simple harmonic oscillator, with damping $\Gamma^2 >  \omega_0^2$ behaves as an underdamped oscillator, whereas, for damping $\Gamma^2 < \omega_0^2$, the system is overdamped. On top of this we will consider the system as an inverted harmonic oscillator, which is obtained when the frequency $\omega_0$ is replaced by $i\,\omega_0$. The motivation for considering the inverted oscillator is to see how the bath can regulate the unstable nature of the inverted harmonic oscillator. 

We start with the reduced density matrix defined in Eq. \ref{rhoSstructure}. We will choose the purification as in \cite{Caceres:2019pgf,Bhattacharyya:2018sbw, Bhattacharyya:2019tsi}
where the authors considered the size of the the ancillary system is the same as the original system. In our case the original subsystem consists of just one oscillator. Hence the purified wavefunction can be parametrized in the following way, 
\begin{equation} \label{pur}
\psi(x,x_{\text{anc}})= \mathcal{N}\exp \left \{-\frac{1}{2} \Big (\alpha x^2+\beta x_{\text{anc}}^2-2\, \tau x x_{\text{anc}}\Big) \right \},
\end{equation}
where $x_{\text{anc}}$ belongs to the ancillary Hilbert-space. $\alpha,\beta $ and $\tau$ are in general complex and yet to be determined. 
The corresponding reduced density matrix after tracing out the auxiliary Hilbert space is
\begin{align}
&\rm {Tr}_{\text{anc}}\ \rho(x, x_{\text{anc}} \, | \, x', x'_{\text{anc}})\\ 
&= \int_{-\infty}^{\infty} dx_{\text{anc}} \, \psi(x,  x_{\text{anc}}) \ \psi^{*} (x', x_{\text{anc}})\\&
= \displaystyle{\mathcal{N}^2\exp\left \{ -\frac{1}{2}\left[\left(\alpha-\frac{\tau^2}{2\, \rm {Re}(\beta)}\right)x^2+\left(\alpha^*-\frac{(\tau^*)^2}{2\, \rm {Re}(\beta)}\right) x'^2\right]+ \left(\frac{|\tau|^2}{2\, \rm {Re}(\beta)} \right ) x x'\right \} },
\end{align}
Comparing with Eq. \ref{rhoSstructure} and \ref{correlator} we get the following,  
\begin{eqnarray} \label{apr}
\alpha - \frac{\tau^2}{2\text{Re} (\beta)} &=& \sigma_P - \frac{4 \sigma^2_{QP} -1+ 4 i \sigma_{QP} }{4 \sigma_Q}, \\
\eta &=& \frac{|\tau|^2}{ 2 \rm {Re}(\beta)}.
\end{eqnarray}
Solving for $\alpha, \rm {Re}(\beta)$ and $\tau$ we get
\begin{equation}
\alpha =\sigma_P, \ \ \rm {Re}(\beta)=2 \sigma_Q, \ \ \tau^2= 4 \sigma^2_{QP} - 1+ 4 i \sigma_{QP}
\end{equation}

We have determined all the parameters in Eq.\ref{pur} in terms of the given parameters as in Eq. \ref{apr}, except for ${\rm Im }(\beta)$. Hence the minimization can be carried over ${\rm Im }(\beta)$ and the minimum value will correspond to the complexity of purification. First, we compute the complexity corresponding to Eq. \ref{pur} with respect to a reference state $\psi_r \sim {\rm exp} \left \{-\frac{1}{2} \omega_r (x^2 +x^2_{\rm anc })\right \} $ as follows:
\begin{equation}
\mathcal{C}(\psi)=\frac{1}{2} \sqrt{ \sum_{i=1}^{2} \left[
     \ln \left (\frac{| \omega_{i}|}{\omega_r} \right)^2 
  + \tan^{-1} \left (-\frac{ \text{Im} ( \omega_{i})}{\text{Re} ( \omega_{i})} \right)^2 
  \right] }  \ ,
\end{equation}
where, 
\begin{align}
\begin{split}
\omega_1=\frac{1}{2}(\alpha+\gamma+\sqrt{(\alpha-\beta)^2+ 4\tau^2}), \ \omega_2=\frac{1}{2}(\alpha+\beta-\sqrt{(\alpha-\beta)^2+ 4\tau^2}),
\ \omega_r=1.
\end{split}
\end{align}
Finally, the complexity of purification is
\begin{equation}
\mathcal{C}_{\rho}=\min_{{\rm Im }(\beta)}\Bigg(\frac{1}{2} \sqrt{ \sum_{i=1}^{2} \left[
     \ln \left (\frac{| \omega_{i}|}{\omega_r} \right)^2 
  + \tan^{-1} \left (-\frac{ \text{Im} ( \omega_{i})}{\text{Re} ( \omega_{i})} \right)^2 
  \right] } \Bigg).
\end{equation}
\begin{figure}[ht]
\begin{center}
\scalebox{0.8}{\includegraphics{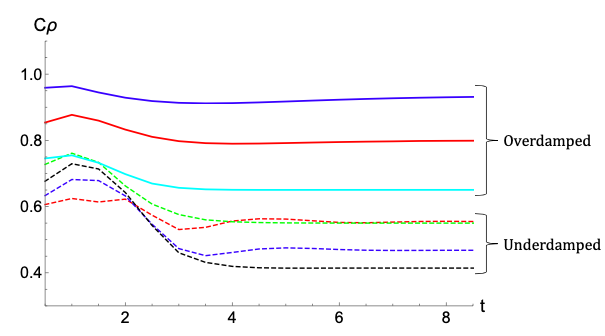}}
\end{center}
\caption{COP vs time. For regular oscillator with: $\omega_0=1$ and various values of the damping \\ $\Gamma = 0.3 \ (\rm red \ dotted), 0.5\ (\rm blue. \ dotted ), 0.7\ (\rm black \ dotted), 0.9\ (\rm green \ dotted),
1.1 \ (\rm cyan), 1.5 \ (\rm red),\\ 2 \ (\rm blue)$.} 
\label{fig:regular}
\end{figure}
Since the reference state considered here is some convenient choice of pure Gaussian state, the complexity of purification will be non-vanishing at $t=0$. 

Below we will investigate the COP for both regular and inverted oscillators:
\begin{equation}
\omega_0=
\begin{cases}
\omega_0 > 0, 
& \mbox{regular oscillator, $\Omega= \sqrt{\omega_0^2- \Gamma^2}$} \cr
i \omega_0, & \mbox{inverted oscillator , $\Omega= i\,\sqrt{\omega_0^2+ \Gamma^2}$}
\end{cases}
\end{equation}
When the oscillator is regular, by changing (increasing) the damping $\Gamma$ we can trace the COP from underdamped ($\Gamma^2<\omega_0^2$) to overdamped ($\Gamma^2 > \omega_0^2$) regime. Fig \ref{fig:regular} displays the COP for a regular oscillator for different values of $\Gamma$. When $\Gamma^2 < \omega_0^2$ the oscillator is underdamped and when $\Gamma^2 > \omega_0^2$ the oscillator is overdamped. The COP quickly saturates for both underdamped and overdamped cases. 

Next we will investigate how the saturation value of the COP depends on the damping $\Gamma$. 
\begin{figure}[ht]
\begin{center}
\scalebox{0.6}{\includegraphics{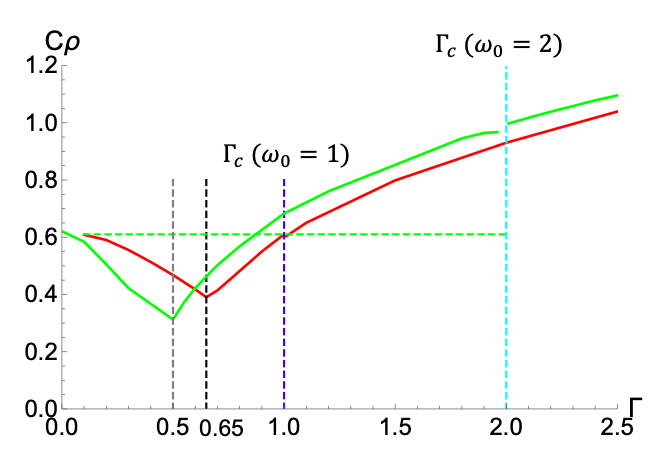}}
\end{center}
\caption{COP vs time. Regular oscillator: the green and magenta line is for $\omega_0=2$ and the red and orange line is for $\omega_0=1$. The critical value of $\Gamma$ appears for $\Gamma_k=0.5$ and $0.65$ respectively. }
\label{fig:regulartransition}
\end{figure}
Fig.\ref{fig:regulartransition} displays the large time behavior (saturation value) of COP  with damping for two different values of the oscillator frequencies, $\omega_0=1$ and $\omega_0=2$.
\begin{figure}[ht]
\begin{center}
\scalebox{0.64}{\includegraphics{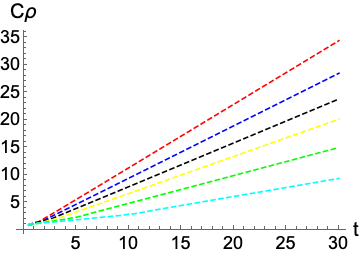}}\ \ 
\scalebox{0.64}{\includegraphics{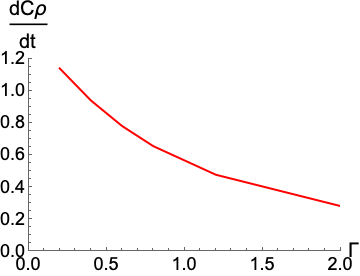}}
\end{center}
\caption{{\bf Left}: COP vs time for the inverted harmonic oscillator with $ \omega_0=i$ and various values of damping $\Gamma = 0.2 {\rm \ (red \ dotted)}, 0.4 {\rm \ (blue \ dotted)},\ 0.6 \ {\rm (black \ dotted)}, \ 0.8 \ {\rm (yellow\ dotted)}, \ 1.2 \\ {\rm (green \ dotted)}, 2.0 \ {\rm (cyan \ dotted)\, respectively;} $ \  {\bf Right}: Rate of change of slope of COP with $\Gamma$.}
\label{fig:inverted}
\end{figure}
To our surprise, we discover the presence of a {\bf kink} for the saturation value of the COP in the underdamped regime at a particular value of the damping, which we will label as $\Gamma_k$. Although, the location of the kink changes with the oscillator frequency $\omega_0$, it always seems to appear at $\Gamma_k <\Gamma_c$. For $\Gamma > \Gamma_k$ the complexity keeps growing as $(a- b e^{-c x})$, with $a=1.5 \ (\omega_0=2), 1.4 \ (\omega_0=1), b= e/2$ and $c= 1/2$. Note that the appearance of the kink does not depend on the choice of the reference frequency $\omega_r$. However, the location changes with the change of the reference frequency. The physical meaning of $\Gamma_k$ is not clear at this point. We would like to explore this further in the future. 

Finally, we consider the harmonic oscillator is in the inverted regime. For that regime we do not get an underdamped case. The system frequency is imaginary for any value damping, hence the system is always overdamped. 
An inverted harmonic oscillator has an unstable fixed point and is a well-known toy model for studying quantum chaos. Recently, the complexity of an isolated inverted harmonic oscillator was studied \cite{Ali:2019zcj,  Bhattacharyya:2020art} and the complexity grows linearly, and the slope of the line plays the role of the Lyapunov exponent when compared with the out of time order correlators.

In our case the oscillator is not isolated and is connected to a bath. However, when the coupling is small one would expect to see a similar unstable behaviour (linear growth of complexity) for the COP. 
Fig. \ref{fig:inverted} illustrates how the COP changes with time for the inverted harmonic oscillator for different values of the damping. The instability of the oscillator is captured by the linear growth as expected. Moreover, we see that as the damping is growing the slope of the COP growth is decreasing, which implies that the instability of the oscillator can be regulated when the interaction between the system and the bath is large. 

\section{Complexity by Operator-State Mapping}
\label{sec:state}
In this section, we will investigate the circuit complexity for our open quantum system by applying operator-state mapping \cite {haqueReducedDensityMatrix}.
Considering explicitly the system's density matrix (Eq.~\ref{rhosystem}), one can associate a state withe $\hat{\rho}_S$ as
\begin{subequations}
\begin{equation}
 \hat{\rho}_S = \int dx dx'~ \rho_S(x,x') ~ | x \rangle \langle x' |  
 \ \  \longleftrightarrow  \ \ 
 | \hat{\rho}_S \rangle = \int dx dx'~ \rho^{\frac{1}{2}}_S(x',x) ~ 
   | x \rangle_{\rm in} \otimes | x' \rangle_{\rm out}  
\end{equation}  
i.e. one has an effective wave function (in this doubled Hilbert space)
\begin{equation}
 \psi(x,x') = \rho^{\frac{1}{2}}_S(x',x) \ . 
\label{effectivepsi}
\end{equation}
\end{subequations}
We will compute the complexity for this effective wave function (Eq.~\ref{effectivepsi}). Using the explicit form of the system's position space density matrix (Eq.~\ref{rhoSstructure}), one obtains
\begin{eqnarray}
 \rho_S(x,x') & = & \sqrt{\frac{\gamma_1 - \eta}{\pi} }
   \exp \left[ - \frac{\gamma}{2} x^2 -\frac{\gamma^*}{2} x'^2 + \eta x x' \right]
 \longleftrightarrow   \nonumber \\
  \rho^{\frac{1}{2}}_S(x,x') & = & \left( \frac{\gamma_1^2 - \eta^2}{\pi^2} \right)^{\frac{1}{4}}
   \exp \left[ - \frac{1}{2} (\gamma + \eta) x^2 - \frac{1}{2} (\gamma^* + \eta) x'^2 
          + \sqrt{2 \eta (\gamma_1 + \eta) } x x' \right]  \ .
\end{eqnarray} 
\begin{figure}[ht]
\begin{center}
\scalebox{0.6}{\includegraphics{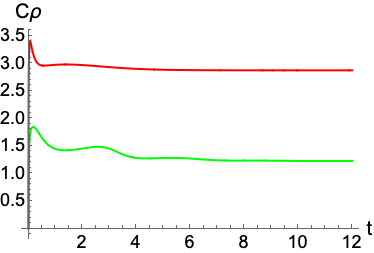} } 
\scalebox{0.6}{\includegraphics{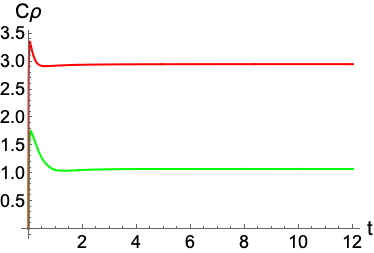} }
\end{center}
\caption{Complexity vs time by operator-state mapping. {\bf Left}: Regular oscillator with $\omega_0=1$ and damping: $ \Gamma = 0.3 \ (\rm green), 1.2 \ (\rm red)$; \ {\bf Right}: Inverted oscillator with $\omega_0=i$ and damping: $\Gamma = 0.3 \ (\rm green), 1.2 \ (\rm red)$.}
\label{fig:stateoperator}
\end{figure}
To proceed, we diagonalize the argument of the exponential; we obtain the effective wave function
\begin{subequations}
\begin{equation}
 \psi(x,x') = \left(\frac{\gamma_1^2 - \eta^2}{\pi^2} \right)^{\frac{1}{4}}
  \exp \left[ - \frac{1}{2} (\xi_1 + E) X^2 - \frac{1}{2} (\xi_1 - E) X'^2 \right],
\end{equation}
where 
\begin{equation}
\kappa^2 = 2 \eta (\gamma_1 + \eta), \ \
 \xi_1 = {\rm Re}[ \gamma + \eta ]  \ \ , \ \ 
 E^2 = \kappa^2 - \xi_2^2 \ \ . \ \
\end{equation}
The new basis $\{X, X'\}$ is related to the old one as follows
\begin{eqnarray} 
 \left( \begin{array}{c}
    X \\ X' 
    \end{array} \right) 
 = \left( \begin{array}{cc}
    u & -v \\ 
    v & u
    \end{array} \right) 
 \left( \begin{array}{c}
    x \\ x' 
\end{array} \right) ,
\end{eqnarray}
\end{subequations}
where
 $ u = \sqrt{ \frac{1}{2} \left( 1 + i \frac{\xi_2}{E} \right)}
 \ , \ 
 v = - \sqrt{ \frac{1}{2} \left( 1 - i \frac{\xi_2}{E} \right)}$.
Interestingly complexity from the state operator mapping for the inverted oscillator 
does not display any linear growth of complexity, hence insensitive to the unstable nature of the inverted oscillator. Moreover, for the regular oscillator we do not see the kink like behaviour as in the COP. The saturation value of the complexity from the operator-state mapping grows with the damping $\Gamma$. This insensitivity is due to the fact that the operator state mapping is only picking out a particular purification and lacks the tracing the minimization procedure as in the COP.


\section{Discussion}
\label{sec:Discussion}
In this work, we explored the complexity of an open quantum system. 
We considered the simplest open quantum system, namely a harmonic oscillator coupled to a one-dimensional (free) bosonic field theory, which served as the ``bath". The reduced density matrix that we get by integrating the bath degrees of freedom served as the target state of our quantum circuit. We applied two known approaches to investigate the complexity of the reduced density matrix, namely the complexity of purification and the complexity, by using the operator state mapping. We considered the system both as regular and inverted oscillators.

 


For both regular and inverted cases, we obtained interesting behaviour for the COP. The COP for regular oscillator saturates for all values of the damping $\Gamma$. However, when we explore the behaviour of COP in the large time limit with varying damping $\Gamma$, we discover the presence of a {\bf kink} for the COP for the underdamped oscillator. The kink appears irrespective of the choice of the reference. This kink represents a new critical value of the damping for open quantum system, which does not show up in any other known physical quantities. The physical meaning of this critical value is not clear at this point and we leave the investigation for a future work.

When the oscillator is inverted, we get a different story. An isolated inverted oscillator is an unstable system. Since our oscillator is connected with a bath, one would expect that the instability of the system will be regulated due to the interaction of the system and the batch. Hence, when the damping due to the bath (string) is low, we expect the system to be more unstable than when the damping is higher. In \cite{Ali:2019zcj}, authors showed that the complexity for an unstable system grows linearly with time and that is exactly what we have observed here; the system behaves more like an isolated inverted oscillator when the damping is low. On the other hand, when the system interacts with the bath strongly, the bath 
can regulate it. More specifically, when the system is in the low damping regime ($\Gamma^2 < \omega_0^2$), the COP {\bf grows linearly} with a larger slope, whereas, when the system is in the high damping regime ($\Gamma^2 >\omega_0^2$), the COP still grows linearly but with much smaller slopes. 

Complexity for the reduced density matrix obtained by using the operator-state mapping seems insensitive compared to COP and displays neither the kink for regular oscillator nor the linear growth for the inverted oscillator. It seems to saturation for both cases for all values of damping. 

In our model, the field satisfies the Dirichlet boundary condition. It would be interesting to consider other boundary conditions (Neumann boundary condition) and the role of the the bath on the system. We leave this investigation for future work. Open quantum systems are more important experimentally, and our findings might be useful in simulating such systems \cite{Breuer:2002pc}.

\section*{Acknowledgements}
We would like to thank Eugene H.~Kim for collaboration in the early stage of this work and Bret Underwood and Chandan Jana for helpful conversations and discussions. S.H. would like to thank the University of Cape Town for funding this project. A.B is supported by  Start-Up Research Grant (SRG/2020/001380) by the Department of Science \& Technology Science and Engineering Research Board (India) and Relevant Research Project grant (58/14/12/2021-BRNS) by the Board Of Research In Nuclear Sciences (BRNS), Department of Atomic Energy, India.


\begin{appendix}

\section{ Periodic Boundary Conditions} \label{A}

We consider the free boson Hamiltonian 
\begin{equation}
 H = \frac{1}{2} \int_0^L \hspace{-0.1in} dx
     \left[ \Pi^2 + (\partial_x \phi)^2 \right]  \ ,
\end{equation}
where $\phi$ satisfies periodic boundary condition $\phi$ satisfies $\phi(x+L) = \phi(x)$.  
To proceed, one introduces the mode expansions
\begin{equation}
 \phi(x) = \frac{1}{\sqrt{L}} \sum_{k} e^{ikx} 
 \frac{1}{\sqrt{2\Omega_k} } ( a^{\phantom \dagger}_{k} + a^{\dagger}_{-k} )
 \ \ , \ \ 
 \Pi(x) = -i \frac{1}{\sqrt{L} } \sum_{k} e^{ikx} 
 \sqrt{\frac{\Omega_k}{2} } ( a^{\phantom \dagger}_k - a^{\dagger}_{-k} )  \ ,
\end{equation}
where $k = (2\pi/L) n$ with $n \in {\bf Z}$, and $\Omega_k = |k|$;
this diagonalizes the Hamiltonian
\begin{equation}
 H = \sum_{k} \Omega_k ( a^{\dagger}_n a^{\phantom \dagger}_n + 1/2 )
 \ .
\end{equation}
One then introduces the dual field $\theta$ such that $\Pi = - \partial_x \theta$;
$\theta$ has the mode expansion
\begin{equation}
 \theta(x) = \frac{1}{\sqrt{L} } \sum_{k} e^{ikx} 
   \frac{\rm sgn(k) }{\sqrt{2 \Omega_k} } ( a^{\phantom \dagger}_k - a^{\dagger}_{-k} )  \ .
\end{equation} 
Finally, one introduces the chiral fields $\phi_R$ and $\phi_L$, such that
\begin{equation}
 \phi = \phi_R + \phi_L  \ \ , \ \ 
 \theta = \phi_R - \phi_L  \ ;
\end{equation}
using the mode expansions for $\phi$ and $\theta$, one obtains
\begin{equation}
 \phi_R(x) = \frac{1}{\sqrt{L}} \sum_{k>0}  \frac{1}{\sqrt{2\Omega_k} } 
  \left( e^{ikx} a^{\phantom \dagger}_{k} + e^{-ikx} a^{\dagger}_{k} \right)
 \ \ , \ \ 
 \phi_L(x) = \frac{1}{\sqrt{L} } \sum_{k>0} \sqrt{\frac{\Omega_k}{2} } 
  \left( e^{-ikx} a^{\phantom \dagger}_{-k} + e^{ikx} a^{\dagger}_{-k} \right)  \ .
\label{chiralmode}
\end{equation} 
We are ultimately interested in $L \rightarrow \infty$ --- 
Eq.~\ref{chiralmode} becomes 
\begin{equation}
 \phi_R(x) = \int_0^{\infty} \hspace{-0.1in} \frac{d\omega}{\sqrt{2\pi} }  \frac{1}{\sqrt{2\omega} } 
  \left( e^{i \omega x} a^{\phantom \dagger}_{\omega} + e^{-i \omega x} a^{\dagger}_{\omega} \right)
 \ \ , \ \ 
 \phi_L(x) =  \int_0^{\infty} \hspace{-0.1in} \frac{d\omega}{\sqrt{2\pi} }  \frac{1}{\sqrt{2\omega} }   
  \left( e^{-i \omega x} a^{\phantom \dagger}_{-\omega} + e^{i \omega x} a^{\dagger}_{-\omega} \right)  \ .
\nonumber
\end{equation} 

\section{Characteristic Function} \label{B}
Consider the following density matrix
\begin{equation} \label{densityCF}
    \rho ( Q, Q') = \mathcal{N} {\rm exp} \left \{ -\frac{1}{2} \left(\alpha Q^2 + \alpha^* Q'^2 -c Q Q'\right)   \right \},
\end{equation}
where $\alpha =(\alpha_1 + i \alpha_2) \in \mathbf{C}$ and $c \in \mathbf{R}$. 
By using this density matrix one can define the $l$'th moment for an arbitrary operator $\mathcal{O}$ as
\begin{equation}
    \langle \mathcal{O}^l \rangle = {\rm tr} (\rho \mathcal{O}^l)
\end{equation}
Now one can define a moment generating (characteristic) function
\begin{equation}
    {\rm CF} (\epsilon, t) \equiv {\rm tr} (\rho e^{-i \epsilon \mathcal{O}}),
\end{equation}
where $\epsilon$ is a real parameter. Moments of all orders can be derived from this characteristic function
\begin{equation}
    \langle \mathcal{O}^l (t) \rangle = \frac{\partial^l}{\partial (i \epsilon)^l} {\rm CF}(\epsilon, t) \ \Big |_{\epsilon=0}
\end{equation}
The generating function for the position and momentum operators can be written in terms of creation and annihilation operators $a$ and $a^{\dagger}$ as follows:
\begin{eqnarray}
 {\rm CF} (\epsilon, t) \equiv {\rm tr} (\rho e^{-i \eta ( \epsilon a +\epsilon^* a^{\dagger})}),
\end{eqnarray}
where $\epsilon \in \mathbf{C}$ and $\eta \in \mathbf{R}$. This is also known as the Wigner characteristic function. The creation and annihilation operators can be written as
\begin{equation}
    a=\frac{1}{\sqrt{2 \Omega}} (\Omega Q+ i P)\ , \ \  a^{\dagger}=\frac{1}{\sqrt{2 \Omega}} (\Omega Q -i P) .
\end{equation}
By using this we can write
\begin{eqnarray} \label{CF3}
 {\rm CF} (\lambda, \mu) \equiv {\rm tr} (\rho \ e^{-i (\lambda Q +\mu P)} ),
\end{eqnarray}
where $\lambda= \eta (\epsilon+\epsilon^*) \sqrt{\Omega/2}$ and $\mu= i \eta (\epsilon - \epsilon^*)/ (\sqrt{2 \Omega})$. By using the commutation relation $( [Q, P]= i )$, the exponent in \ref{CF3} can be expanded as 
\begin{eqnarray}
  e^{-i (\lambda Q +\mu P)} &=& e^{-i \lambda Q} e^{-i\mu P} e^{\frac{1}{2}i\lambda \mu} \nonumber \\ 
  &=& e^{-\frac{1}{2} i\mu P} e^{ \frac{1}{2}i\mu P} e^{-i \lambda Q} e^{- \frac{1}{2} i\mu P} e^{- \frac{1}{2} i\mu P} e^{\frac{1}{2}i\lambda \mu} \nonumber \\
  &=& e^{- \frac{1}{2} i\mu P} e^{ -i\lambda (Q+\frac{\mu}{2})} e^{ -\frac{1}{2} i\mu P} e^{ \frac{1}{2} i\lambda \mu} \nonumber \\
  &=& e^{ -\frac{1}{2} i\mu P} e^{-i\lambda Q} e^{ \frac{1}{2} i\mu P}
\end{eqnarray}
 Now we will evaluate the trace in the position representation, where
 \begin{eqnarray}
  Q|Q'\rangle &=& Q'|Q'\rangle \\
  \langle Q'|Q''\rangle &=&\delta (Q'-Q'')\\
  \int_{-\infty}^{\infty} dQ'|Q'\rangle \langle Q'| & =& 1
 \end{eqnarray}
 Furthermore,
 \begin{eqnarray}
 e^{ \frac{1}{2} i\mu P} |Q'\rangle &=& {\Big|} Q' - \frac{\mu}{2}{\Big \rangle} \\
\langle Q'| \ e^{ \frac{1}{2} i\mu P}&=& {\Big \langle} Q' + \frac{\mu}{2}{\Big |}
 \end{eqnarray}
Using these we get from \ref{CF3}
\begin{eqnarray}
 {\rm CF} (\lambda, \mu) & = & {\rm tr} \  e^{ -\frac{1}{2} i\mu P} \rho \  e^{ -\frac{1}{2} i\mu P}  e^{-i\lambda Q} \\
 & =& \int  e^{-i\lambda Q'} {\Big \langle} Q' -\frac{\mu}{2} |\rho| Q'+\frac{\mu}{2} {\Big \rangle} dQ'
\end{eqnarray}
Now plugging the density matrix (Eq. \ref {densityCF}) in the above expression and performing the gaussian integrals we get
\begin{equation}
     {\rm CF} (\lambda, \mu) = {\rm exp} \left\{ -\frac{A}{2} \lambda^2 - \frac{A'}{2} \mu^2 +B \mu \lambda \right\}
\end{equation}
where the parameters $A, A'$ and $B$ are
\begin{eqnarray}
A &=& \frac{1}{2 (\alpha_1 -c)} \\
A' &=& \frac{|\alpha|^2-c^2}{2 (\alpha_1 -c)}\\
B &=& -\frac{ \alpha_2}{2 (\alpha_1-c)}.
\end{eqnarray}
These parameters are related to the correlators as follows: 
\begin{eqnarray}
 \langle Q Q \rangle &=& - \frac{\partial^2}{\partial \lambda^2} {\rm CF (\lambda, \mu)} {\Big |}_{\mu=\lambda=0} = A \\
 \langle P P \rangle &=& - \frac{\partial^2}{\partial \mu^2} {\rm CF (\lambda, \mu)} {\Big |}_{\mu=\lambda=0} = A' \\
\frac{1}{2} \langle QP+PQ \rangle &=& - \frac{\partial^2}{\partial \mu \partial \lambda} {\rm CF (\lambda, \mu)} {\Big |}_{\mu=\lambda=0} = B
\end{eqnarray}

\section{Correlation Functions} \label{C}
The central objects are the system correlators 
$\langle \hat{O}_{\alpha}(t_1) \hat{O}_{\beta}(t_2) \rangle$, 
where $\hat{O}_{\alpha,\beta}$ = $Q$ or $P$.
A key role is played by the $Q_p$ correlator, $\langle Q_p(t_1) Q_p(t_2) \rangle$.  
Explicitly, 
\begin{equation}
 Q_P(t) = \int_0^{\infty} \hspace{-0.1in} \frac{d\omega}{\sqrt{2\pi} } 
  \frac{1}{\sqrt{2\omega} } \left[
     f(\omega) e^{-i\omega t} a^{\phantom \dagger}_{\omega} 
  + f(-\omega) e^{i\omega t} a^{\dagger}_{\omega} \right]  
 \ \ {\rm with} \ \ 
   f(\omega) = \frac{i\omega 2 \sqrt{2\Gamma} }
                     {\omega^2 + i2 \Gamma \omega - \omega_0^2}  \ .
\end{equation}
Using that the bath is taken to be initially in its ground state, one obtains
\begin{equation}
 \langle Q_p(t_1) Q_p(t_2) \rangle = \frac{2\Gamma}{\pi} \int_0^{\infty} \hspace{-0.1in} d\omega~
 e^{-i\omega(t_1 - t_2) } \frac{\omega}{ (\omega^2 - \omega_0^2 )^2 + (2\Gamma \omega)^2 } \ .
\nonumber
\end{equation} \label{xxcorrelator}
Carrying out the integral(s), one obtains 
\begin{eqnarray}
 \langle Q_p(t_1) Q_p(t_2) \rangle & = & I_1 + i I_2 
\end{eqnarray}
where
\begin{eqnarray}
  I_1 & = & - \frac{1}{4 \pi \sqrt{\Gamma^2-\omega_0^2}} \left ( 
    2 \left[ \sin (\tau \omega_+) {\rm Si} (\tau \omega_+) 
  - \sin (\tau \omega^*_+) {\rm Si} (\tau \omega^*_+) \right]  \right.
 \nonumber \\  & & \hspace{-0.0in}  \left. 
 + \cos (\tau \omega_+) \left[ {\rm Ci} (\tau \omega_+) + {\rm Ci} (- \tau \omega_+) \right] 
 -  \cos (\tau \omega^*_+) \left[ {\rm Ci} (\tau \omega^*_+) + {\rm Ci} (- \tau \omega^*_+) \right] 
   \right )  \ ,
 \nonumber \\
 I_2 & = &  \frac{1}{i (4 \sqrt{\Gamma^2-\omega_0^2})} \left( \sin (\omega_+ \tau) -\sin (\omega_+^* \tau)\right) \ .
\end{eqnarray}
In the above equation, $\omega_+ =\sqrt{\omega_0^2 + \ 2\Gamma (-\Gamma +\sqrt{\Gamma^2- \omega_0^2}) }$, $\omega_+^* = \sqrt{\omega_0^2 -  2\Gamma (-\Gamma +\sqrt{\Gamma^2- \omega_0^2}) }$,
$\tau = t_1 - t_2$, and ${\rm Si}(z)$ (${\rm Ci}(z)$) is the sine-integral (cosine-integral) function \cite{nist}. 
We also compute the other correlators:
\begin{eqnarray}  \label{xpcorrelator} 
 \langle \dot Q_p(t_1) Q_p(t_2) \rangle & = & J_1 + i J_2,
 \end{eqnarray}
 where
 \begin{eqnarray}
 J_1 & = & - \frac{1}{4 \sqrt{2} \pi \sqrt{\Gamma^2-\omega_0^2}} \left\{ 
 2  \left[ \omega_+ \cos (\tau \omega_+) {\rm Si} (\tau \omega_+) 
  -  \omega^*_+ \cos (\tau \omega^*_+) {\rm Si} (\tau \omega^*_+) \right]  \right.
 \nonumber \\  & & \hspace{-0.0in}  \left. 
 -  \omega_+\sin (\tau \omega_+) \left[ {\rm Ci} (\tau \omega_+) + {\rm Ci} (- \tau \omega_+) \right] 
 +  \omega^*_+\sin (\tau \omega^*_+) \left[ {\rm Ci} (\tau \omega^*_+) + {\rm Ci} (- \tau \omega^*_+) \right] 
   \right\}  \ ,
 \nonumber \\
J_2 & = & \frac{1}{i (4 \sqrt{\Gamma^2-\omega_0^2})} \left( \omega_+ \cos (\omega_+ \tau) - \omega_+^* \cos (\omega_+^* \tau) \right).
\end{eqnarray}
Similarly we get
\begin{eqnarray}  \label{ppcorrelator} 
 \langle \dot Q_p(t_1) \dot Q_p(t_2) \rangle & = & K_1 + i K_2,
 \end{eqnarray}
 where
 \begin{eqnarray}
 K_1 & = & - \frac{1}{8 \pi \sqrt{\Gamma^2-\omega_0^2}} \left\{ 
 2  \left[ \omega_+^2 \sin (\tau \omega_+) {\rm Si} (\tau \omega_+) 
  -  \omega^{*2}_+ \sin (\tau \omega^*_+) {\rm Si} (\tau \omega^*_+) \right]  \right.
 \nonumber \\  & & \hspace{-0.0in}  \left. 
 + \omega_+^2 \cos (\tau \omega_+) \left[ {\rm Ci} (\tau \omega_+) + {\rm Ci} (- \tau \omega_+) \right] 
 -  \omega^{*2}_+\cos (\tau \omega^*_+) \left[ {\rm Ci} (\tau \omega^*_+) + {\rm Ci} (- \tau \omega^*_+) \right] 
   \right\}  \ ,
 \nonumber \\
K_2 & =& \frac{1}{i (4 \sqrt{\Gamma^2-\omega_0^2})} \left ( \omega_+^2 \sin (\omega_+ \tau) - \omega_+^{*2} \sin (\omega_+^* \tau) \right )
\end{eqnarray}
All combinations of 
\begin{eqnarray}  
 \langle Q(t_1) Q_p(t_2) \rangle & = & 0
 \end{eqnarray}
and 
\begin{equation} 
 \langle Q Q \rangle  =  \frac{1}{2 \Omega}, \ \ \ \langle \dot Q \dot Q \rangle  =  \frac{1}{2 \Omega}.
 \end{equation}
 Finally, note that to carry out the computations discussed in the paper, we need equal time correlators as mentioned in Eq.~\ref{correlator}. But in the limit $t_1 \rightarrow t_2$, they are typically divergent. To regularize the integral, we have used a cutoff $\tau_c.$


\end{appendix}


\bibliographystyle{utphysmodb}
\bibliography{refs}

\providecommand{\href}[2]{#2}\begingroup\raggedright\begin{thebibliography}{10}

\bibitem{CALDEIRA1983374}
A.~Caldeira and A.~Leggett,  {\em Quantum tunnelling in a dissipative system},
  Annals of Physics {\bf 149} (1983), no.~2, 374 -- 456.

\bibitem{PhysRevLett.46.211}
A.~O. Caldeira and A.~J. Leggett,  {\em Influence of Dissipation on Quantum
  Tunneling in Macroscopic Systems}, Phys. Rev. Lett. {\bf 46} (Jan, 1981)
  211--214.

\bibitem{zurekbath}
W.~G. Unruh and W.~H. Zurek,  {\em Reduction of a wave packet in quantum
  Brownian motion}, Phys. Rev. D {\bf 40} (Aug, 1989) 1071--1094.

\bibitem{Ali:2019zcj}
T.~Ali, A.~Bhattacharyya, S.~S. Haque, E.~H. Kim, N.~Moynihan and J.~Murugan,
  {\em {Chaos and Complexity in Quantum Mechanics}}, Phys. Rev. D {\bf 101}
  (2020), no.~2, 026021 [\href{http://www.arXiv.org/abs/1905.13534}{{\tt
  1905.13534}}].

\bibitem{Jefferson}
R.~Jefferson and R.~C. Myers,  {\em {Circuit complexity in quantum field
  theory}}, JHEP {\bf 10} (2017) 107
[\href{http://www.arXiv.org/abs/1707.08570}{{\tt 1707.08570}}].

\bibitem{Chapman:2017rqy}
S.~Chapman, M.~P. Heller, H.~Marrochio and F.~Pastawski,  {\em {Toward a
  Definition of Complexity for Quantum Field Theory States}}, Phys. Rev. Lett.
  {\bf 120} (2018), no.~12, 121602
  [\href{http://www.arXiv.org/abs/1707.08582}{{\tt 1707.08582}}].

\bibitem{Caputa:2017yrh}
P.~Caputa, N.~Kundu, M.~Miyaji, T.~Takayanagi and K.~Watanabe,  {\em {Liouville
  Action as Path-Integral Complexity: From Continuous Tensor Networks to
  AdS/CFT}}, JHEP {\bf 11} (2017) 097
  [\href{http://www.arXiv.org/abs/1706.07056}{{\tt 1706.07056}}].

\bibitem{me1}
T.~Ali, A.~Bhattacharyya, S.~Shajidul~Haque, E.~H. Kim and N.~Moynihan,  {\em
  {Time Evolution of Complexity: A Critique of Three Methods}}, JHEP {\bf 04}
  (2019) 087 [\href{http://www.arXiv.org/abs/1810.02734}{{\tt 1810.02734}}].

\bibitem{Bhattacharyya:2018bbv}
A.~Bhattacharyya, A.~Shekar and A.~Sinha,  {\em {Circuit complexity in
  interacting QFTs and RG flows}}, JHEP {\bf 10} (2018) 140
  [\href{http://www.arXiv.org/abs/1808.03105}{{\tt 1808.03105}}].

\bibitem{Hackl:2018ptj}
L.~Hackl and R.~C. Myers,  {\em {Circuit complexity for free fermions}}, JHEP
  {\bf 07} (2018) 139 [\href{http://www.arXiv.org/abs/1803.10638}{{\tt
  1803.10638}}].

\bibitem{Khan:2018rzm}
R.~Khan, C.~Krishnan and S.~Sharma,  {\em {Circuit Complexity in Fermionic
  Field Theory}}, Phys. Rev. D {\bf 98} (2018), no.~12, 126001
  [\href{http://www.arXiv.org/abs/1801.07620}{{\tt 1801.07620}}].

\bibitem{Camargo:2018eof}
H.~A. Camargo, P.~Caputa, D.~Das, M.~P. Heller and R.~Jefferson,  {\em
  {Complexity as a novel probe of quantum quenches: universal scalings and
  purifications}}, Phys. Rev. Lett. {\bf 122} (2019), no.~8, 081601
  [\href{http://www.arXiv.org/abs/1807.07075}{{\tt 1807.07075}}].

\bibitem{Ali:2018aon}
T.~Ali, A.~Bhattacharyya, S.~Shajidul~Haque, E.~H. Kim and N.~Moynihan,  {\em
  {Post-Quench Evolution of Complexity and Entanglement in a Topological
  System}}, Phys. Lett. B {\bf 811} (2020) 135919
  [\href{http://www.arXiv.org/abs/1811.05985}{{\tt 1811.05985}}].

\bibitem{Bhattacharyya:2018wym}
A.~Bhattacharyya, P.~Caputa, S.~R. Das, N.~Kundu, M.~Miyaji and T.~Takayanagi,
  {\em {Path-Integral Complexity for Perturbed CFTs}}, JHEP {\bf 07} (2018) 086
  [\href{http://www.arXiv.org/abs/1804.01999}{{\tt 1804.01999}}].

\bibitem{Caputa:2018kdj}
P.~Caputa and J.~M. Magan,  {\em {Quantum Computation as Gravity}}, Phys. Rev.
  Lett. {\bf 122} (2019), no.~23, 231302
  [\href{http://www.arXiv.org/abs/1807.04422}{{\tt 1807.04422}}].

\bibitem{Bhattacharyya:2019kvj}
A.~Bhattacharyya, P.~Nandy and A.~Sinha,  {\em {Renormalized Circuit
  Complexity}}, Phys. Rev. Lett. {\bf 124} (2020), no.~10, 101602
[\href{http://www.arXiv.org/abs/1907.08223}{{\tt 1907.08223}}].

\bibitem{Caputa:2020mgb}
P.~Caputa and I.~MacCormack,  {\em {Geometry and Complexity of Path Integrals
  in Inhomogeneous CFTs}}, \href{http://www.arXiv.org/abs/2004.04698}{{\tt
  2004.04698}}.

\bibitem{Flory:2020eot}
M.~Flory and M.~P. Heller,  {\em {Complexity and Conformal Field Theory}},
  \href{http://www.arXiv.org/abs/2005.02415}{{\tt 2005.02415}}.

\bibitem{Erdmenger:2020sup}
J.~Erdmenger, M.~Gerbershagen and A.-L. Weigel,  {\em {Complexity measures from
  geometric actions on Virasoro and Kac-Moody orbits}},
  \href{http://www.arXiv.org/abs/2004.03619}{{\tt 2004.03619}}.

\bibitem{cosmology1}
A.~Bhattacharyya, S.~Das, S.~S. Haque and B.~Underwood,  {\em Cosmological
  complexity}, Physical Review D {\bf 101} (May, 2020).

\bibitem{cosmology2}
A.~Bhattacharyya, S.~Das, S.~S. Haque and B.~Underwood,  {\em Rise of
  cosmological complexity: Saturation of growth and chaos}, Physical Review
  Research {\bf 2} (Aug, 2020).

\bibitem{DiGiulio:2020hlz}
G.~Di~Giulio and E.~Tonni,  {\em {Complexity of mixed Gaussian states from
  Fisher information geometry}},
  \href{http://www.arXiv.org/abs/2006.00921}{{\tt 2006.00921}}.

\bibitem{Caceres:2019pgf}
E.~Caceres, S.~Chapman, J.~D. Couch, J.~P. Hernandez, R.~C. Myers and S.-M.
  Ruan,  {\em {Complexity of Mixed States in QFT and Holography}}, JHEP {\bf
  03} (2020) 012 [\href{http://www.arXiv.org/abs/1909.10557}{{\tt
  1909.10557}}].

\bibitem{Susskind:2020gnl}
L.~Susskind and Y.~Zhao,  {\em {Complexity and Momentum}},
  \href{http://www.arXiv.org/abs/2006.03019}{{\tt 2006.03019}}.

\bibitem{Chen:2020nlj}
B.~Chen, B.~Czech and Z.-z. Wang,  {\em {Cutoff Dependence and Complexity of
  the CFT$_2$ Ground State}}, \href{http://www.arXiv.org/abs/2004.11377}{{\tt
  2004.11377}}.

\bibitem{Czech:2017ryf}
B.~Czech,  {\em {Einstein Equations from Varying Complexity}}, Phys. Rev. Lett.
  {\bf 120} (2018), no.~3, 031601
  [\href{http://www.arXiv.org/abs/1706.00965}{{\tt 1706.00965}}].

\bibitem{Camargo:2019isp}
H.~A. Camargo, M.~P. Heller, R.~Jefferson and J.~Knaute,  {\em {Path integral
  optimization as circuit complexity}}, Phys. Rev. Lett. {\bf 123} (2019),
  no.~1, 011601 [\href{http://www.arXiv.org/abs/1904.02713}{{\tt 1904.02713}}].

\bibitem{Chapman:2018hou}
S.~Chapman, J.~Eisert, L.~Hackl, M.~P. Heller, R.~Jefferson, H.~Marrochio and
  R.~C. Myers,  {\em {Complexity and entanglement for thermofield double
  states}}, SciPost Phys. {\bf 6} (2019), no.~3, 034
  [\href{http://www.arXiv.org/abs/1810.05151}{{\tt 1810.05151}}].

\bibitem{Chapman:2019clq}
S.~Chapman and H.~Z. Chen,  {\em {Complexity for Charged Thermofield Double
  States}}, \href{http://www.arXiv.org/abs/1910.07508}{{\tt 1910.07508}}.

\bibitem{Doroudiani:2019llj}
M.~Doroudiani, A.~Naseh and R.~Pirmoradian,  {\em {Complexity for Charged
  Thermofield Double States}}, JHEP {\bf 01} (2020) 120
  [\href{http://www.arXiv.org/abs/1910.08806}{{\tt 1910.08806}}].

\bibitem{Geng:2019yxo}
H.~Geng,  {\em {$T\bar{T}$ Deformation and the Complexity=Volume Conjecture}},
  Fortsch. Phys. {\bf 68} (2020), no.~7, 2000036
  [\href{http://www.arXiv.org/abs/1910.08082}{{\tt 1910.08082}}].

\bibitem{Guo:2020dsi}
M.~Guo, Z.-Y. Fan, J.~Jiang, X.~Liu and B.~Chen,  {\em {Circuit complexity for
  generalized coherent states in thermal field dynamics}}, Phys. Rev. {\bf
  D101} (2020), no.~12, 126007
[\href{http://www.arXiv.org/abs/2004.00344}{{\tt 2004.00344}}].

\bibitem{Haque:2021hyw}
S.~S. Haque, C.~Jana and B.~Underwood,  {\em {Operator Complexity for
  Continuous Variable Systems}},
  \href{http://www.arXiv.org/abs/2110.08356}{{\tt 2110.08356}}.

\bibitem{Haque:2021kdm}
S.~S. Haque, C.~Jana and B.~Underwood,  {\em {Saturation of Thermal Complexity
  of Purification}}, \href{http://www.arXiv.org/abs/2107.08969}{{\tt
  2107.08969}}.

\bibitem{Chapman:2021jbh}
S.~Chapman and G.~Policastro,  {\em {Quantum Computational Complexity -- From
  Quantum Information to Black Holes and Back}},
  \href{http://www.arXiv.org/abs/2110.14672}{{\tt 2110.14672}}.

\bibitem{Bhattacharyya:2021cwf}
A.~Bhattacharyya,  {\em {Circuit complexity and (some of) its applications}},
  Int. J. Mod. Phys. E {\bf 30} (2021), no.~07, 2130005.

\bibitem{susskind1}
L.~Susskind,  {\em {Computational Complexity and Black Hole Horizons}},
  Fortsch. Phys. {\bf 64} (2016) 24--43
  [\href{http://www.arXiv.org/abs/1403.5695}{{\tt 1403.5695}}], [Addendum:
  Fortsch.Phys. 64, 44--48 (2016)].

\bibitem{susskind2}
L.~Susskind,  {\em {Entanglement is not enough}}, Fortsch. Phys. {\bf 64}
  (2016) 49--71 [\href{http://www.arXiv.org/abs/1411.0690}{{\tt 1411.0690}}].

\bibitem{Agn2019}
C.~A. Agon, M.~Headrick and B.~Swingle,  {\em Subsystem complexity and
  holography}, Journal of High Energy Physics {\bf 2019} (Feb, 2019).

\bibitem{Camargo:2020yfv}
H.~A. Camargo, L.~Hackl, M.~P. Heller, A.~Jahn, T.~Takayanagi and B.~Windt,
  {\em {Entanglement and Complexity of Purification in (1+1)-dimensional free
  Conformal Field Theories}}, \href{http://www.arXiv.org/abs/2009.11881}{{\tt
  2009.11881}}.

\bibitem{Balasubramanian:2019wgd}
V.~Balasubramanian, M.~Decross, A.~Kar and O.~Parrikar,  {\em {Quantum
  Complexity of Time Evolution with Chaotic Hamiltonians}}, JHEP {\bf 01}
  (2020) 134 [\href{http://www.arXiv.org/abs/1905.05765}{{\tt 1905.05765}}].

\bibitem{Bhattacharyya:2019txx}
A.~Bhattacharyya, W.~Chemissany, S.~Shajidul~Haque and B.~Yan,  {\em Towards
  the Web of Quantum Chaos Diagnostics},
  \href{http://www.arXiv.org/abs/1909.01894}{{\tt 1909.01894}}.

\bibitem{Ryuchaos}
J.~Kudler-Flam, L.~Nie and S.~Ryu,  {\em {Conformal field theory and the web of
  quantum chaos diagnostics}}, JHEP {\bf 01} (2020) 175
  [\href{http://www.arXiv.org/abs/1910.14575}{{\tt 1910.14575}}].

\bibitem{Bhattacharyya:2020art}
A.~Bhattacharyya, W.~Chemissany, S.~S. Haque, J.~Murugan and B.~Yan,  {\em {The
  Multi-faceted Inverted Harmonic Oscillator: Chaos and Complexity}}, SciPost
  Phys. Core {\bf 4} (2021) 002
  [\href{http://www.arXiv.org/abs/2007.01232}{{\tt 2007.01232}}].

\bibitem{Yang:2019iav}
R.-Q. Yang and K.-Y. Kim,  {\em {Time evolution of the complexity in chaotic
  systems: a concrete example}}, JHEP {\bf 05} (2020) 045
  [\href{http://www.arXiv.org/abs/1906.02052}{{\tt 1906.02052}}].

\bibitem{Yang:2019udi}
R.-Q. Yang, Y.-S. An, C.~Niu, C.-Y. Zhang and K.-Y. Kim,  {\em {To be
  unitary-invariant or not?: a simple but non-trivial proposal for the
  complexity between states in quantum mechanics/field theory}},
  \href{http://www.arXiv.org/abs/1906.02063}{{\tt 1906.02063}}.

\bibitem{CHOI1975285}
M.-D. Choi,  {\em Completely positive linear maps on complex matrices}, Linear
  Algebra and its Applications {\bf 10} (1975), no.~3, 285 -- 290.

\bibitem{JAMIOLKOWSKI1972275}
A.~Jamiolkowski,  {\em Linear transformations which preserve trace and positive
  semidefiniteness of operators}, Reports on Mathematical Physics {\bf 3}
  (1972), no.~4, 275 -- 278.

\bibitem{haqueReducedDensityMatrix}
A.~Bhattacharyya, S.~S. Haque and E.~H. Kim,  {\em {Complexity from the Reduced
  Density Matrix: a new Diagnostic for Chaos}}, JHEP {\bf 10} (2021) 028
  [\href{http://www.arXiv.org/abs/2011.04705}{{\tt 2011.04705}}].

\bibitem{buttiker}
M.~B\"uttiker,  {\em Scattering theory of current and intensity noise
  correlations in conductors and wave guides}, Phys. Rev. B {\bf 46} (Nov,
  1992) 12485--12507.

\bibitem{Bhattacharyya:2018sbw}
A.~Bhattacharyya, T.~Takayanagi and K.~Umemoto,  {\em {Entanglement of
  Purification in Free Scalar Field Theories}}, JHEP {\bf 04} (2018) 132
  [\href{http://www.arXiv.org/abs/1802.09545}{{\tt 1802.09545}}].

\bibitem{Bhattacharyya:2019tsi}
A.~Bhattacharyya, A.~Jahn, T.~Takayanagi and K.~Umemoto,  {\em {Entanglement of
  Purification in Many Body Systems and Symmetry Breaking}}, Phys. Rev. Lett.
  {\bf 122} (2019), no.~20, 201601
  [\href{http://www.arXiv.org/abs/1902.02369}{{\tt 1902.02369}}].

\bibitem{Breuer:2002pc}
H.~P. Breuer and F.~Petruccione, {\em {The theory of open quantum systems}}.
\newblock 2002.

\bibitem{nist}
F.~Olver, D.~Lozier, R.~Boisvert and C.~Clark, {\em NIST Handbook of
  Mathematical Functions}.
\newblock 01, 2010.

\end{thebibliography}\endgroup

\end{document}